\documentclass[aps,twocolumn,showpacs]{revtex4}
\usepackage{amsmath}
\usepackage{epsfig}
\usepackage{booktabs} %调整表格线与上下内容的间隔
\usepackage{multirow}

\begin{document}

\title{The study of double-charm and double-strange tetraquarks}

\author{Yuheng Wu$^1$}\email[E-mail: ]{191002007@njnu.edu.cn}
\author{Xin Jin$^1$}\email[E-mail: ]{181002005@njnu.edu.cn}
\author{Runxin Liu$^1$}\email[E-mail: ]{201002011@njnu.edu.cn}
\author{Hongxia Huang$^1$}\email[E-mail: ]{hxhuang@njnu.edu.cn (Corresponding author)}
\author{Jialun Ping$^1$}\email[E-mail: ]{jlping@njnu.edu.cn (Corresponding author)}
\affiliation{$^1$Department of Physics, Nanjing Normal University, Nanjing, Jiangsu 210097, China}

\begin{abstract}
In the framework of the quark delocalization color screening model (QDCSM), we systematically investigate the double-charm and double-strange tetraquark systems $cc\bar{s}\bar{s}$ with two structures: meson-meson and diquark-antidiquark. The bound-state calculation shows that there is no any bound state in present work. However, by applying a stabilization calculation and coupling all channels of both two structures, two new resonance states with $IJ^{P}=00^{+}$ are obtained. One is more possible to be a molecular resonance state with a mass and width around $4256$ MeV and $60$ MeV, respectively; another one is more likely to be a compact resonance state with a mass and width around $4308$ MeV and $19$ MeV, respectively. Although no significant signals were observed in present experiment at the Belle collaboration, there is still some structures around $4.3$ GeV in the distributions of $M_{D^{+}_{s}D^{+}_{s}}$ and $M_{D^{*+}_{s}D^{*+}_{s}}$. We suggest that the experiment can be further tested with a larger amount of data.
\end{abstract}

\maketitle

\setcounter{totalnumber}{5}

\section{\label{sec:introduction}Introduction}
Over the past decades, dozens of exotic hadron states have been reported in experiments worldwide. These states provide us an ideal platform to
deepen our understanding of the non-perturbative quantum chromodynamics (QCD). Although none of the exotic states is now definitely confirmed by experiment,
more and more theoretical work have been done to investigate the exotic states. Searching for new exotic states is one of the most significant research topics in the hadron physics.

The hidden-charm and hidden-strange tetraquark, which is composed of $cs\bar{c}\bar{s}$, is one type of the exotic states. In 2009, the CDF Collaboration reported the $X(4140)$ in the $J/\psi\phi$ invariant mass distribution~\cite{CDF}.
Later, the $X(4140)$ was successively observed by other experiments, such as Belle~\cite{Bell}, CMS~\cite{CMS}, D0~\cite{D0}, BABAR~\cite{BABAR}, and LHCb~\cite{LHCb-CCSS} collaborations. In 2010, a narrow resonance $X(4350)$ was observed in the $\gamma\gamma\longrightarrow J/\psi\phi$ process by the Belle collaboration~\cite{Bell}. In 2016, LHCb collaboration reported four exotic states $X(4140)$, $X(4274)$, $X(4500)$, and $X(4700)$ from the amplitude analysis of the $B^{+}\rightarrow J/\psi\phi K^{+}$ decays~\cite{LHCb1,LHCb2}. In 2017, the $X(4274)$ was reported by the CDF collaboration in the process of $B^{+}\rightarrow J/\psi\phi K^{+}$~\cite{CDF2}. In 2020, the LHCb collaboration reported a new state $\chi_{c0}(3930)$, which is just below the $D_{s}^{-}D_{s}^{+}$ threshold~\cite{LHCb_X3930.1,LHCb_X3930.2}. In 2021, two new hadron states $X(4685)$ and $X(4630)$ were reported with the help of the improved full amplitude analysis of the $B^+\longrightarrow J/\psi \phi K^+$ decay by the LHCb collaboration\cite{LHCb-CCSS-2021}.

These experimental developments aroused great interests in studying the hidden-charm and hidden-strange tetraquarks. In Ref.~\cite{XJL}, Liu {\em et al.} investigated the tetraquark composed of $cs\bar{c}\bar{s}$, and found that the $X(4350)$, $X(4500)$, and $X(4700)$ could be explained as the compact tetraquarks with $IJ^{P}=00^{+}$ and the $X(4274)$ was explained as a compact tetraquark with $IJ=01^{+}$. In the framework of the chiral quark model, Yang {\em et al.} found that the $X(4274)$ could be the $cs\bar{c}\bar{s}$ tetraquark with $J^{PC}=1^{++}$ and the $X(4350)$ could be assigned as a candidate of the compact tetraquark with $J^{PC}=0^{++}$. For the $X(4700)$, it is explained as the $2S$ radial excited tetraquark with $J^{PC}=0^{++}$~\cite{YFY}. In Ref.~\cite{CHX}, the $cs\bar{c}\bar{s}$ tetraquarks were studied within the diquark-antidiquark configuration by using the QCD sum rule. They claimed that the $X(4140)$ and $X(4274)$ could be assigned as the $S-$wave $cs\bar{c}\bar{s}$ tetraquarks with opposite color structures and both the $X(4500)$ and $X(4700)$ were the $D-$wave $cs\bar{c}\bar{s}$ tetraquarks with opposite color structures too. More results and discussions are given in Refs.~\cite{JWu,WZG1,QFL,PGO,LM,DCR1,ESS}.

Recently, the BESIII collaboration reported a new structure $Z_{cs}(3985)^{-}$ near the $D_{s}^{-}D^{*0}/D_{s}^{*-}D^{0}$ thresholds in the processes of $e^{+}e^{-} \rightarrow K^{+}(D_{s}^{-}D^{*0}+D_{s}^{*-}D^{0})$. The mass and width of this state are ($3982.5^{+1.8}_{-2.6}\pm2.1$) Mev and ($12.8^{+5.3}_{-4.4}\pm3.0$) Mev, respectively~\cite{BESIII:2020qkh}. From the production mode, the minimum quark component of $Z_{cs}(3985)^{-}$ is $cs\bar{c}\bar{u}$, so it is the first candidate of the charged hidden-charm tetraquark state with strangeness. Later, the LHCb collaboration reported two new states $Z_{cs}(4000)^{+}$ and $Z_{cs}(4220)^{+}$ with the quark content of $cu\bar{c}\bar{s}$ decaying to the $J/\psi K^{+}$~\cite{LHCb-CCSS-2021}. The decay widths of these two states are $\Gamma=131\pm15\pm26$ MeV and $\Gamma=233\pm52^{+97}_{-73}$  MeV, respectively. Obviously, the masses of $Z_{cs}(3985)^{-}$ and $Z_{cs}(4000)^{+}$ are close, but the decay widths are largely different.
These observations immediately stimulated a lot of theoretical studies of the open-strange and hidden-charm tetraquarks ~\cite{JinX,Sun:2020hjw,Cao:2020cfx,Du:2020vwb,Wang:2020kej,Yang:2020nrt,Azizi:2020zyq,Wang:2020htx,Simonov:2020ozp,Wang:2020iqt,Chen:2020yvq,Meng:2020ihj,Wan:2020oxt,Rossi:2020ezg,Yan:2021tcp,Wang:2020rcx}
Some works showed that $Z_{cs}(3985)^{-}$ and $Z_{cs}(4000)^{+}$ are the same states~\cite{Ortega:2021enc,Giron:2021sla}, while some indicated that they are not the same states~\cite{Yang:2020nrt,Meng:2021rdg,Chen:2021erj,Shi:2021jyr}.

Inspired by the study progress of the hidden-charm and hidden-strange tetraquarks, and the open-strange and hidden-charm tetraquarks, it is natural to investigate the existence of the open-charm and open-strange tetraquarks. Very recently, the Belle collaboration searched for the double-heavy tetraquark state candidates $X_{cc\bar{s}\bar{s}}$ decaying to $D^{+}_{s}D^{+}_{s}$ and $D^{*+}_{s}D^{*+}_{s}$, but no significant
signals were observed~\cite{Belle2021}. In Ref.~\cite{Yanggang11}, the double-charm and double-strange tetraquarks $cc\bar{s}\bar{s}$ were studied within the chiral quark model, and some resonance states with $IJ^{P}=00^{+}$ and $02^{+}$ were obtained. In Ref.~\cite{R. Molina}, the $D_{s}^{*+}D_{s}^{*+}$ systems with $IJ^{P}=00^{+}$ and $02^{+}$ were studied in a coupled channel unitary approach, but no any bound state was found because of the strong repulsion.

As is commonly believed that QCD is the fundamental theory of the strong interaction.
However, the low energy physics of QCD, such as hadron structure, hadron-hadron interactions, and the structure
of multiquark systems, is much harder to calculate directly from QCD. The quark delocalization color
screening model (QDCSM), which was developed in 1990s with the aim of explaining the similarities
between nuclear and molecular forces~\cite{QDCSM0}, is one of the effective approaches for studying the multiquark systems. Two new ingredients were introduced: quark
delocalization (to enlarge the model variational space to take into account the mutual distortion
or the internal excitations of nucleons in the course of their interactions, the distortion of wave
functions in the existence of other nucleons is also considered in the quark-meson-coupling model~\cite{QMC})
and color screening (assuming the quark-quark interaction dependent on quark states aimed to take into
account the QCD effect which has not yet been included in the two-body confinement and effective one gluon
exchange). This model has been applied to the study of the dibaryos~\cite{Ping1,Huang2}, pentaquarks~\cite{Pc_Huang} and some tetraquark systems~\cite{XJL}.
It is also interesting to extend this model to study the open-charm and open-strange tetraquarks. As the first step, we investigate the existence of the double-charm and double-strange tetraquarks $cc\bar{s}\bar{s}$ in this work. Different structures and the effect of the channel-coupling are considered.

The structure of this paper is as follows. Section II gives a brief description of the quark model and wave functions. Section III is devoted to the numerical results and discussions. The summary is shown in the last section.

\section{MODEL AND WAVE FUNCTIONS}

\subsection{The quark delocalization color screening model (QDCSM)}
In this paper, we use the quark delocalization color screening model (QDCSM) to investigate the $cc\bar{s}\bar{s}$ tetraquark system. The details of the QDCSM can be found in the Refs.~\cite{QDCSM0,Huang1,QDCSM1}. Here we just present the Hamiltonian of the model.
\begin{widetext}
\begin{eqnarray}
H & = & \sum_{i=1}^4 \left( m_i+\frac{p_i^2}{2m_i}\right)-T_{CM} +\sum_{j>i=1}^4
\left(V_{ij}^{CON}+V_{ij}^{OGE}+V_{ij}^{OBE} \right),\\
V_{ij}^{CON} & = & \left \{
\begin{array}{ll}
-a_{c}\boldsymbol {\mathbf{\lambda}}^c_{i}\cdot
\boldsymbol{\mathbf{ \lambda}}^c_{j}~\left(r_{ij}^2+a^{0}_{ij}\right),&
   \mbox{if \textit{i},\textit{j} in the same baron orbit}\\
-a_{c}\boldsymbol {\mathbf{\lambda}}^c_{i}\cdot
\boldsymbol{\mathbf{\lambda}}^c_{j}~(
\frac{1-e^-\mu_{ij}\mathbf{r}_{ij}^2}{\mu_{ij}}+a^0_{ij}),& \mbox{otherwise}
\end{array}
\right.\label{QDCSM-vc}\\
V^{OGE}_{ij} & = & \frac{1}{4}\alpha_s \boldsymbol{\lambda}^{c}_i \cdot
\boldsymbol{\lambda}^{c}_j
\left[\frac{1}{r_{ij}}-\frac{\pi}{2}\delta(\boldsymbol{r}_{ij})(\frac{1}{m^2_i}+\frac{1}{m^2_j}
+\frac{4\boldsymbol{\sigma}_i\cdot\boldsymbol{\sigma}_j}{3m_im_j})%-\frac{3}{4m_im_jr^3_{ij}}
\right] \label{sala-vG} \\
%S_{ij}\right] \label{sala-vG} \\
%V^{OBE}_{ij} & = & V_{\pi}( \boldsymbol{r}_{ij})\sum_{a=1}^3\lambda
%_{i}^{a}\cdot \lambda
%_{j}^{a}+V_{K}(\boldsymbol{r}_{ij})\sum_{a=4}^7\lambda
%_{i}^{a}\cdot \lambda _{j}^{a}
%+V_{\eta}(\boldsymbol{r}_{ij})\left[\left(\lambda _{i}^{8}\cdot
%\lambda _{j}^{8}\right)\cos\theta_P-(\lambda _{i}^{0}\cdot
%\lambda_{j}^{0}) \sin\theta_P\right] \label{sala-Vchi1} \\
V^{OBE}_{ij} & = & V_{\eta}(\boldsymbol{r}_{ij})\left[\left(\lambda _{i}^{8}\cdot
\lambda _{j}^{8}\right)\cos\theta_P-(\lambda _{i}^{0}\cdot
\lambda_{j}^{0}) \sin\theta_P\right] \label{sala-Vchi1} \\
%V_{\chi}(\boldsymbol{r}_{ij}) & = & {\frac{g_{ch}^{2}}{{4\pi
%}}}{\frac{m_{\chi}^{2}}{{\
%12m_{i}m_{j}}}}{\frac{\Lambda _{\chi}^{2}}{{\Lambda _{\chi}^{2}-m_{\chi}^{2}}}}%
%m_{\chi} \left\{(\boldsymbol{\sigma}_{i}\cdot
%\boldsymbol{\sigma}_{j})
%\left[ Y(m_{\chi}\,r_{ij})-{\frac{\Lambda_{\chi}^{3}}{m_{\chi}^{3}}}%
%Y(\Lambda _{\chi}\,r_{ij})\right] \right.\nonumber \\
%&& \left. +\left[H(m_{\chi}
%r_{ij})-\frac{\Lambda_{\chi}^3}{m_{\chi}^3}
%H(\Lambda_{\chi} r_{ij})\right] S_{ij} \right\}, ~~~~~~\chi=\pi, K, \eta, \\
V_{\eta}(\boldsymbol{r}_{ij}) & = & {\frac{g_{ch}^{2}}{{4\pi
}}}{\frac{m_{\eta}^{2}}{{\
12m_{i}m_{j}}}}{\frac{\Lambda _{\eta}^{2}}{{\Lambda _{\eta}^{2}-m_{\eta}^{2}}}}%
m_{\eta} \left\{(\boldsymbol{\sigma}_{i}\cdot
\boldsymbol{\sigma}_{j})
\left[ Y(m_{\eta}\,r_{ij})-{\frac{\Lambda_{\eta}^{3}}{m_{\eta}^{3}}}%
Y(\Lambda _{\eta}\,r_{ij})\right] \right\}.\nonumber \\
%&& \left. +\left[H(m_{\chi}
%r_{ij})-\frac{\Lambda_{\chi}^3}{m_{\chi}^3}
%H(\Lambda_{\chi} r_{ij})\right] S_{ij} \right\}, ~~~~~~\chi=\pi, K, \eta, \\
%S_{ij}&=&\left\{ 3\frac{(\boldsymbol{\sigma}_i
%\cdot\boldsymbol{r}_{ij}) (\boldsymbol{\sigma}_j\cdot
%\boldsymbol{r}_{ij})}{r_{ij}^2}-\boldsymbol{\sigma}_i \cdot
%\boldsymbol{\sigma}_j\right\},\\
%H(x)&=&(1+3/x+3/x^{2})Y(x),~~~~~~
Y(x) &=& e^{-x}/x. \label{sala-vchi2}
\end{eqnarray}
\end{widetext}
where the $T_{CM}$ is the kinetic energy of the center of mass; $V_{ij}^{CON}$ and $V_{ij}^{OGE}$ are the confinement and one-gluon-exchange interactions, respectively; $V_{ij}^{OBE}$ is the Goldstone-boson exchange interaction. In Eq.(2), the $\mu_{ij}$ is the color screening parameter, which is determined by fitting the
deuteron properties, $NN$ scattering phase shifts, and $N\Lambda$ and $N\Sigma$ scattering phase shifts, with
$\mu_{uu}=0.45~fm^{-2}$, $\mu_{us}=0.19~fm^{-2}$, $\mu_{ss}=0.08~fm^{-2}$, satisfying the relation,
$\mu_{us}^2=\mu_{uu}\mu_{ss}$. When extending to the heavy-quark sector, we found that the dependence of the parameter $\mu_{cc}$ is not very significant in the calculation of the $P_{c}$ states~\cite{Pc_huang1} by taking it from $0.0001~fm^{-2}$ to $0.01~fm^{-2}$.
So here we take $\mu_{cc}=0.01~fm^{-2}$. Then $\mu_{sc}$ and $\mu_{uc}$ are obtained by the relation $\mu^{2}=\mu_{ss}\mu_{cc} $ and $\mu^{2}=\mu_{uu}\mu_{cc}$, respectively. Here, we focus on the $S-$wave $cc\bar{s}\bar{s}$ states, so the tensor force interaction is not included. The $V_{ij}^{OGE}$ can be briefly written as Eq.(3), where the $\alpha_{s}$ is the quark-gluon coupling constant. For the $V_{ij}^{OBE}$, there are no $\pi$ and $K$ meson exchange in the $cc\bar{s}\bar{s}$ system. So, we only use the $\eta$ exchange here. In Eq.(5), the $Y(x)=e^{-x}/x$ is the standard Yukawa function; $g_{ch}$ is the coupling constant for chiral field, which is determined from the $NN\pi$ coupling constant through
\begin{equation}
\frac{g_{ch}^2}{4\pi}=\left(\frac{3}{5}\right)^2\frac{g_{\pi NN}^2}{4\pi}\frac{m_{u,d}^2}{m_N^2}
\end{equation}
The other symbols in the above expressions have their usual meanings. All model parameters, which are determined by fitting the meson spectrum, are from the work of $c\bar{c}s\bar{s}$ system~\cite{XJL}.

The quark delocalization in QDCSM is achieved by writing the single-particle orbital wave function as a linear combination of the left and right Gaussian functions, the single particle orbital wave functions used in the ordinary quark cluster model,
\begin{eqnarray}
\psi_{\alpha}(\mathbf{s}_i ,\epsilon) & = & \left(
\phi_{\alpha}(\mathbf{s}_i)
+ \epsilon \phi_{\alpha}(-\mathbf{s}_i)\right) /N(\epsilon), \nonumber \\
\psi_{\beta}(-\mathbf{s}_i ,\epsilon) & = &
\left(\phi_{\beta}(-\mathbf{s}_i)
+ \epsilon \phi_{\beta}(\mathbf{s}_i)\right) /N(\epsilon), \nonumber \\
N(\epsilon) & = & \sqrt{1+\epsilon^2+2\epsilon e^{-s_i^2/4b^2}}, \label{1q} \\
\phi_{\alpha}(\mathbf{s}_i) & = & \left( \frac{1}{\pi b^2}
\right)^{3/4}
   e^{-\frac{1}{2b^2} (\mathbf{r}_{\alpha} - \mathbf{s}_i/2)^2}, \nonumber \\
\phi_{\beta}(-\mathbf{s}_i) & = & \left( \frac{1}{\pi b^2}
\right)^{3/4}
   e^{-\frac{1}{2b^2} (\mathbf{r}_{\beta} + \mathbf{s}_i/2)^2}. \nonumber
\end{eqnarray}
Where the $\mathbf{s}_i$, $i=1,2,...,n$ are the generating coordinates, which are introduced to expand the relative motion wavefunction~\cite{QDCSM1}. The $\epsilon(\mathbf{s}_i)$ is determined variationally by the dynamics of the multi-quark system itself rather than an adjustable one, which can make the system choose the  favorable configuration in the interacting process.

\subsection{Wave function}
The resonating group method (RGM)~\cite{RGM}, a well established method for studying a bound-state or a scattering problem, is used to calculate the energy of all these states in this work. The wave function of the four-quark system is of the form
\begin{equation}
\Psi_{4q}=\mathcal{A}\sum_{L}[[\Psi_{A}\Psi_{B}]^{[\sigma]IS}\bigotimes\chi_{L}(\boldsymbol{\mathbf{R}})]^{J},
\end{equation}
The symbol ${\cal A }$ is the anti-symmetrization operator. $[\sigma]=[222]$ gives the total color symmetry,  except that all other symbols have the usual meanings. $\Psi_{A}$ and $\Psi_{B}$ are the 2-quark cluster wave functions,
\begin{equation}
\Psi_{A}=(\frac{1}{2\pi b^{2}})^{3/4}e^{-\boldsymbol{\mathbf{\rho}}^{2}_{A}/(4b^{2})}\eta _{I_{A}S_{A}}\chi_{A}^{c},
\end{equation}

\begin{equation}
\Psi_{B}=(\frac{1}{2\pi b^{2}})^{3/4}e^{-\boldsymbol{\mathbf{\rho}}_{B}^{2}/(4b^{2})}\eta _{I_{B}S_{B}}\chi_{B}^{c},
\end{equation}
Where $\eta _{I_{A}S_{A}}$/$\eta _{I_{B}S_{B}}$ represent the multiplied wave functions of flavor and spin of the cluster A/B. $\chi_{A}^{c}$/$\chi_{B}^{c}$ are the internal color wave functions of cluster A/B, and the Jacobi coordinates are shown as:
\begin{eqnarray}
\boldsymbol{\mathbf{\rho}}_{A}=\boldsymbol{\mathbf{r}}_{1}-\boldsymbol{\mathbf{r}}_{2},
\qquad
\boldsymbol{\mathbf{\rho}}_{B}=\boldsymbol{\mathbf{r}}_{3}-\boldsymbol{\mathbf{r}}_{4},\nonumber \\
\boldsymbol{\mathbf{R}}_{A}=\frac{1}{2}(\boldsymbol{\mathbf{r}}_{1}+\boldsymbol{\mathbf{r}}_{2}),
\qquad
\boldsymbol{\mathbf{R}}_{B}=\frac{1}{2}(\boldsymbol{\mathbf{r}}_{3}+\boldsymbol{\mathbf{r}}_{4}),\nonumber\\
\boldsymbol{\mathbf{R}}=\boldsymbol{\mathbf{R}}_{A}-\boldsymbol{\mathbf{R}}_{B},
\qquad
\boldsymbol{\mathbf{R}}_{C}=\frac{1}{2}(\vec{R}_{A}+\vec{R}_{B}),
\end{eqnarray}

From the variational principle, after variation with respect to the relative motion wave function $\chi(\boldsymbol{\mathbf{R}})=\sum_{L}\chi_{L}(\boldsymbol{\mathbf{R}})$, one obtains the RGM equation
\begin{equation}
\int H(\boldsymbol{\mathbf{R}},\boldsymbol{\mathbf{R'}})\chi(\boldsymbol{\mathbf{R'}})d\boldsymbol{\mathbf{R'}}=E\int N(\boldsymbol{\mathbf{R}},\boldsymbol{\mathbf{R'}})\chi(\boldsymbol{\mathbf{R'}})d\boldsymbol{\mathbf{R'}}
\end{equation}
where $H(\boldsymbol{\mathbf{R}},\boldsymbol{\mathbf{R'}})$ and  $N(\boldsymbol{\mathbf{R}},\boldsymbol{\mathbf{R'}})$ are Hamiltonian and norm kernels, By solving the RGM equation, we can get the energies $E$ and the wave functions. In fact, it is not convenient to work with the RGM expressions. Then, we use the gussian bases to expand the relative motion wave function $\chi(\boldsymbol{\mathbf{R}})$, respectively.
\begin{eqnarray}
& & \chi_{L}(\boldsymbol{R}) = \frac{1}{\sqrt{4\pi}}\sum_{L}(\frac{1}{\pi b^2})^{\frac{3}{4}} \sum_{i} C_{i,L}  \nonumber \\
&& ~~~~\times  \int e^{-\frac{1}{2}(\boldsymbol{R}-\boldsymbol{S}_{i})^{2}/b^{2}} Y^{L}(\hat{\boldsymbol{S}_{i}})d\hat{\boldsymbol{S}_{i}}. ~~~~~
\end{eqnarray}
where $\boldsymbol{S_{i}}$ is the separation of two reference centers, and plays the role of the generator coordinate in the model; $C_{i,L}$ is the expansion coefficient.
After the inclusion of the center of mass motion,
\begin{equation}
\Phi_{C}(\boldsymbol{R}_{C})=(\frac{4}{\pi b^{2}})^{3/4} e^{-2}\boldsymbol{R}^{2}_{C}/b^{2},
\end{equation}
the ansatz, Eq.(5), can be rewritten as
\begin{eqnarray}
\Psi_{4q}= {\cal A} \sum_{i,L} C_{i,L}
\int\frac{d\Omega_{S_{i}}}{\sqrt{4\pi}}\prod^{2}_{\alpha=1}\phi_{\alpha}(\boldsymbol{S_{i}})\prod^{4}_{\beta=3}\phi_{\beta}(-\boldsymbol{S_{i}})\nonumber\\
\times[[\eta_{I_{A}S_{A}}\eta_{I_{B}S_{B}}]^{IS}Y^{L}(\boldsymbol{S_{i}})]^{J}[\chi_{A}^{c}\chi_{B}^{c}]^{[\sigma]},
\end{eqnarray}
where $\phi_{\alpha}(\boldsymbol{S_{i}})$ and $\phi_{\beta}(-\boldsymbol{S_{i}})$ are the single-particle orbital wave functions with different reference centers:
\begin{eqnarray}
\phi_{\alpha}(\boldsymbol{S_{i}})=(\frac{1}{\pi b^{2}})^{3/4}e^{-\frac{1}{2}(\boldsymbol{r}_{\alpha}-\boldsymbol{S_{i}}/2)^{2}/b^{2}},\nonumber\\
\phi_{\beta}(\boldsymbol{-S_{i}})=(\frac{1}{\pi b^{2}})^{3/4}e^{-\frac{1}{2}(\boldsymbol{r}_{\beta}+\boldsymbol{S_{i}}/2)^{2}/b^{2}},
\end{eqnarray}
With the reformulated ansatz, Eq.(15), the RGM equation (12) becomes an algebraic eigenvalue equation:
\begin{equation}
\sum_{j,L}C_{j,L}H^{L,L'}_{i,j}=E\sum_{j}C_{j,L'}N^{L'}_{i,j},
\end{equation}
where $N^{L'}_{i,j}$ and $H^{L,L'}_{i,j}$ are the wave function (15) overlaps and Hamiltonian matrix elements (without the summation over $L'$ ), respectively. By solving the generalized eigen problem, we can obtain the energies of the 4-quark systems $E$ and corresponding expansion coefficient $C_{j,L}$. Finally, the relative motion wave function between two clusters can be obtained by substituting the $C_{j,L}$ into Eq. (13). The flavor, spin and color wave functions are constructed in the following part.

In this work, the flavor wave function for the tetraquark system we investigate is $cc\bar{s}\bar{s}$. Different structures are obtained according to different coupling sequences. For the meson-meson structure, the coupling sequence is
\begin{equation}
\chi_{m}^{f1}=c\bar{s}-c\bar{s}
\end{equation}
For the diquark-antidiquark structure, the coupling sequence is
\begin{equation}
\chi_{d}^{f1}=cc-\bar{s}\bar{s}
\end{equation}
Note that this coupling sequence should match the orbital coupling sequence. For the coupling sequence $c\bar{s}-c\bar{s}$, the orbital coordinates are defined as Eq. (11); for the coupling sequence $cc-\bar{s}\bar{s}$, the orbital coordinates in Eq. (11) change to
\begin{eqnarray}
\boldsymbol{\mathbf{\rho}}_{A}=\boldsymbol{\mathbf{r}}_{1}-\boldsymbol{\mathbf{r}}_{3},
\qquad
\boldsymbol{\mathbf{\rho}}_{B}=\boldsymbol{\mathbf{r}}_{2}-\boldsymbol{\mathbf{r}}_{4},\nonumber \\
\boldsymbol{\mathbf{R}}_{A}=\frac{1}{2}(\boldsymbol{\mathbf{r}}_{1}+\boldsymbol{\mathbf{r}}_{3}),
\qquad
\boldsymbol{\mathbf{R}}_{B}=\frac{1}{2}(\boldsymbol{\mathbf{r}}_{2}+\boldsymbol{\mathbf{r}}_{4}),\nonumber\\
\end{eqnarray}
So, both structures have the same anti-symmetrization operator:
\begin{equation}
{\cal A } = 1-P_{13}-P_{24}+P_{13}P_{24}\nonumber
\end{equation}

For the spin wave functions of meson-meson structure, firstly, we construct the two-body spin wave functions as:
\begin{eqnarray}
\chi^{1}_{\sigma_{11}} &=& \alpha\alpha,~~~~\chi^{2}_{\sigma_{10}} = \sqrt{\frac{1}{2}}(\alpha\beta+\beta\alpha), \nonumber \\
~~~~\chi^{3}_{\sigma_{1-1}} &=& \beta\beta,~~~~\chi^{4}_{\sigma_{00}} = \sqrt{\frac{1}{2}}(\alpha\beta-\beta\alpha).
\end{eqnarray}
Then the spin wave functions of meson-meson structure can be obtained by coupling the wave functions of two clusters:
\begin{eqnarray}
\psi^{1}_{0} &=& \chi^{4}_{\sigma_{00}}\chi^{4}_{\sigma_{00}}\nonumber\\
\psi^{2}_{0} &=& \sqrt{\frac{1}{3}}(\chi^{1}_{\sigma_{11}}\chi^{3}_{\sigma_{1-1}}-\chi^{2}_{\sigma_{10}}\chi^{2}_{\sigma_{10}}+\chi^{3}_{\sigma_{1-1}}\chi^{1}_{\sigma_{11}})
\nonumber\\
\psi^{3}_{1} &=& \chi^{4}_{\sigma_{00}}\chi^{1}_{\sigma_{11}}\nonumber\\
\psi^{4}_{1} &=& \chi^{1}_{\sigma_{11}}\chi^{4}_{\sigma_{00}}\nonumber\\
\psi^{5}_{1} &=& \sqrt{\frac{1}{2}}(\chi^{1}_{\sigma_{11}}\chi^{2}_{\sigma_{10}}-\chi^{2}_{\sigma_{10}}\chi^{1}_{\sigma_{11}})\nonumber\\
\psi^{6}_{1} &=& \chi^{1}_{\sigma_{11}}\chi^{1}_{\sigma_{11}}
\end{eqnarray}
For the diquark-antidiquark structure, the spin wave functions are the same as the meson-meson structure.

Finally, for the color wave function, two structures are much different. We construct the color wave function in the same way as the spin wave functions. The color wave function for a $q\bar{q}$ cluster is:
\begin{eqnarray}
\chi^{1}_{[111]} &=& \sqrt{\frac{1}{3}}(r\bar{r}+g\bar{g}+b\bar{b}).
\end{eqnarray}
Then the color wave function of the meson-meson structure is:
\begin{eqnarray}
\psi^{c_{1}} &=& \chi^{1}_{[111]}\chi^{1}_{[111]}.
\end{eqnarray}

However, the situation is even more complicated for diquark-antidiquark structure. We construct the color wave function of the $qq$ and $\bar{q}\bar{q}$
clusters firstly.
The color wave functions of the $qq$ clusters are:
\begin{eqnarray}
\chi^{1}_{[2]} &=& rr,~~~\chi^{2}_{[2]} = \frac{1}{\sqrt{2}}(rg+gr),~~~\chi^{3}_{[2]} = gg,  \nonumber \\
\chi^{4}_{[2]} &=& \frac{1}{\sqrt{2}}(rb+br),~~~\chi^{5}_{[2]} = \frac{1}{\sqrt{2}}(gb+bg), \nonumber \\
\chi^{6}_{[2]} &=& bb,~~~\chi^{7}_{[11]} = \frac{1}{\sqrt{2}}(rg-gr), \nonumber \\
\chi^{8}_{[11]}&=& \frac{1}{\sqrt{2}}(rb-br),~~~\chi^{9}_{[11]} = \frac{1}{\sqrt{2}}(gb-bg).
\end{eqnarray}
and the color wave functions of the $\bar{q}\bar{q}$ clusters are:
\begin{eqnarray}
\chi^{1}_{[22]} &=& \bar{r}\bar{r},~~~\chi^{2}_{[22]} = -\frac{1}{\sqrt{2}}(\bar{r}\bar{g}+\bar{g}\bar{r}),~~~~\chi^{3}_{[22]} = \bar{g}\bar{g},  \nonumber \\
\chi^{4}_{[22]} &=& \frac{1}{\sqrt{22}}(\bar{r}\bar{b}+\bar{b}\bar{r}),~\chi^{5}_{[22]} = -\frac{1}{\sqrt{2}}(\bar{g}\bar{b}+\bar{b}\bar{g}),\nonumber \\
\chi^{6}_{[22]} &=& \bar{b}\bar{b},~~~\chi^{7}_{[211]} = \frac{1}{\sqrt{2}}(\bar{r}\bar{g}-\bar{g}\bar{r}),\nonumber \\
\chi^{8}_{[211]}&=& -\frac{1}{\sqrt{2}}(\bar{r}\bar{b}-\bar{b}\bar{r}),~~~\chi^{9}_{[211]} =  \frac{1}{\sqrt{2}}(\bar{g}\bar{b}-\bar{b}\bar{g}).
\end{eqnarray}
Then the color wave functions of $qq-\bar{q}\bar{q}$ structure are shown as:
\begin{eqnarray}
\psi^{c_{1}} &=& \sqrt{\frac{1}{6}}[\chi^{1}_{[2]}\chi^{1}_{[22]}-\chi^{2}_{[2]}\chi^{2}_{[22]}+\chi^{3}_{[2]}\chi^{3}_{[22]}  \nonumber \\
&& +\chi^{4}_{[2]}\chi^{4}_{[22]}-\chi^{5}_{[2]}\chi^{5}_{[22]}+\chi^{6}_{[2]}\chi^{6}_{[22]}],  \nonumber \\
\psi^{c_{2}} &=& \sqrt{\frac{1}{3}}\left[\chi^{7}_{[11]}\chi^{7}_{[211]}-\chi^{8}_{[11]}\chi^{8}_{[211]}+\chi^{9}_{[11]}\chi^{9}_{[211]}\right]
\end{eqnarray}

Finally, we can acquire the total wave functions by substituting the wave functions of the orbital, the spin, the flavor and the color parts into the Eq. (8) according to the definite quantum number of the system.

\section{Result and discussion}
In this work, we investigate the double-charm and double-strange tetraquark system $cc\bar{s}\bar{s}$ in the framework of QDCSM. Two structures: meson-meson and diquark-antiqiquark structures, as well as the channel-coupling of the two configurations are considered. Since we are focus on the $S-$wave states, the orbital angular momentum is set to be zero. The spin quantum number of the $cc\bar{s}\bar{s}$ system can be $0,~1$ and $2$, so the total angular momentum can be $J=0,~1$ and $2$ for this system. The isospin of the $c$ or $s$ quark is zero. In this way, the quantum number of the $cc\bar{s}\bar{s}$ tetraquark system can be $IJ^{P}=00^{+}, 01^{+}$ and $02^{+}$.
The energy of the $cc\bar{s}\bar{s}$ tetraquark systems for both the meson-meson and diquark-antidiquark structures, as well as the channel coupling of these two structures are listed in Table~\ref{energy}, where the $E_{sc}$ is the energy of every single channel, $E_{cc}$ shows the energy by channel coupling of one certain configuration, and $E_{mix}$ is the lowest energy of the system by coupling all channels of both two configurations.

\begin{table}[ht]
\caption{\label{energy} The energies of the $cc\bar{s}\bar{s}$ system. }
\begin{tabular}{ccccccccc} \hline\hline
$IJ^{P}$ &~~~$Channel$ &~~~$Threshold$ &~~~$E_{sc}$ &~~$E_{cc}$ &~~~$E_{mix}$~~~~\\
 $00^{+}$   &~~     $D_{s}^{+}D_{s}^{+}$      &~~ 3936       &~~~ 3942      &~~~~ 3942      &   3939    \\% \hline
            &~~     $D_{s}^{*+}D_{s}^{*+}$      &~~ 4224       &~~~ 4228   \\
            &~~    $(cc)_{\bar{3}}(\bar{s}\bar{s})_{3}$      &~~        &~~~ 4370    &~~~~ 4312\\
            &~~     $(cc)_{\bar{6}}(\bar{s}\bar{s})_{6}$      &~~        &~~~ 4412    \\
 \\
 $01^{+}$   &~~     $D_{s}^{+}D_{s}^{*+}$      &~~ 4080       &~~~ 4086      &~~~~ 4086      &   4083    \\
            &~~     $(cc)_{\bar{3}}(\bar{s}\bar{s})_{3}$      &~~        &~~~ 4390         &~~~~ 4390 \\
 \\
 $02^{+}$   &~~     $D_{s}^{*+}D_{s}^{*+}$      &~~ 4224       &~~~ 4230      &~~~~ 4230      &   4224    \\
            &~~     $(cc)_{\bar{3}}(\bar{s}\bar{s})_{3}$      &~~        &~~~ 4428          &~~~~ 4428\\

\hline\hline
\end{tabular}
\end{table}

For the $IJ^{P}=00^{+}$ system, there are four channels, which are $D_{s}^{+}D_{s}^{+}$, $D_{s}^{*+}D_{s}^{*+}$, and two diquark-antidiquark channels with the color configurations $(\bar{3}\times3)$ and $(\bar{6}\times6)$. For the meson-meson structure, the energies of both the $D_{s}^{+}D_{s}^{+}$ and $D_{s}^{*+}D_{s}^{*+}$ channels are above the corresponding threshold, which means that neither of them are bound state. The lowest energy is almost unchanged after the channel-coupling calculation, which means that the effect of the channel-coupling here is very small. This is mainly due to the large mass gap between the $D_{s}^{+}D_{s}^{+}$ and $D_{s}^{*+}D_{s}^{*+}$ channels. For the diquark-antidiquark structure,
it is obviously that the energies of both two channels are much higher than those of the meson-meson structure. Although the energy is pushed down about $60$ MeV by coupling these two channels, it is still higher than the meson-meson structure. By coupling all channels of both two structures, the lowest energy is still above the threshold of the $D_{s}^{+}D_{s}^{+}$, which indicates that there is no bound state for the $IJ^{P}=00^{+}$ system.

For the $IJ^{P}=01^{+}$ system, the threshold of $D_{s}^{+}D_{s}^{*+}$ is $4080$ MeV and the energies of the two structures are $4086$ MeV, $4390$ MeV, respectively, both of which are higher than the threshold of $D_{s}^{+}D_{s}^{*+}$. Then, channel-coupling of two structures have been performed and the energy $E_{mix}=4083$ MeV is obtained, which is still higher than the threshold. Therefore, no any bound state is obtained for the $IJ^{P}=01^{+}$ system.

For the $IJ^{P}=02^{+}$ system, the case is similar with the one of the $IJ^{P}=01^{+}$ system. The energy of each single channel is above the threshold of the $D_{s}^{*+}D_{s}^{*+}$. The channel-coupling cannot help too much. So there is no bound state for the $IJ^{P}=02^{+}$ system, either.

Although there is no any bound state for the $IJ^{P}=00^{+}, 01^{+}$ and $02^{+}$ systems, some resonance states are still possible. Since the colorful subclusters diquark and antidiquark cannot fall apart directly due to the color confinement, it is possible for them to be resonance states.
To find out if there is any resonance state,  stabilization method, also named as a real scaling method, which has proven to be a valuable tool for estimating the energies of the metastable states of electron-atom, electron-molecule, and atom-diatom complexes~\cite{real_method1}, is employed to find the genuine resonances. In this method, with the increase of the distance between two clusters, the continuum state will fall off towards its threshold, while a resonance state will tend to be stable. In this situation, the resonance line acts as an avoid-crossing structure and it will appear repeatedly with the increment of the distance between two clusters. Then the resonance line corresponds to the energy of the resonance state. This method has been successfully applied to the pentaquark systems~\cite{real_method2,real_method3}, the fully-heavy tetraquark systems~\cite{real_method4}, and the $c\bar{c}s\bar{s}$ tetraquark systems~\cite{XJL}. Here, we calculate the energy eigenvalues of the $cc\bar{s}\bar{s}$ tetraquark systems by taking the value of the distance ($S$) between two clusters from $4.5$ fm to $8.5$ fm to see if there is any stable state.
The results of the $cc\bar{s}\bar{s}$ tetraquark systems with $IJ^{P}=00^{+}, 01^{+}$ and $02^{+}$ are shown in Figs.1-3, respectively.

For the $cc\bar{s}\bar{s}$ system with $IJ^{P}=00^{+}$ in Fig. 1, it is clearly that the first two horizontal lines locate at the corresponding physical threshold of two channels $D_{s}^{+}D_{s}^{+}$ and $D_{s}^{*+}D_{s}^{*+}$. We mark them with red lines. Another two lines around $4256$ MeV and $4308$ MeV are stable with the variation of the distance between two clusters, so both of them are on behalf of resonance states. We mark them with blue lines. Besides, we also calculate the component of each channel for these two resonance states. For the resonance state with the mass of $4256$ MeV, the proportion of the channels with the meson-meson structure is about $90\%$, while the one of the diquark-antidiqaurk structure is about $10\%$. It indicates that this state is more likely to be the molecular structure.  For the resonance state with the mass of $4308$ MeV, the proportion of the channels with the meson-meson structure is about $40\%$, while the one of the diquark-antidiqaurk structure is about $60\%$. It means that this resonance state is more inclined to be the compact structure rather than the molecular structure.

Moreover, the decay width of this resonance state can be calculated by the formula~\cite{real_method1}:
\begin{eqnarray}
\Gamma &=& 4 V(S) \frac{ \sqrt{(k_r \times k_c)}} {\lvert k_r-k_c \rvert}.
\end{eqnarray}
where the $V(S)$ is the minimal energy difference between the resonance state and the scattering state, $k_r$ and $k_c$ stand for the slope of the resonance state and the scattering state, respectively. Then, we obtain the width of these two resonance state is about $60$ MeV and $19$ MeV, respectively.

The results of the $cc\bar{s}\bar{s}$ systems with $IJ^{P}=01^{+}$ and $02^{+}$ are shown in Fig. 2 and Fig. 3, respectively. The first horizontal line in two figures represents the threshold of the $D_{s}^{+}D_{s}^{*+}$ and $D_{s}^{*+}D_{s}^{*+}$, respectively. It is obvious that with the increase of the distance between two clusters, the energy of the continuum state falls off towards its threshold. So there is no any resonance state for the $cc\bar{s}\bar{s}$ systems with $IJ^{P}=01^{+}$ and $02^{+}$.

\begin{figure}[ht]
\begin{center}
\epsfxsize=3.39in \epsfbox{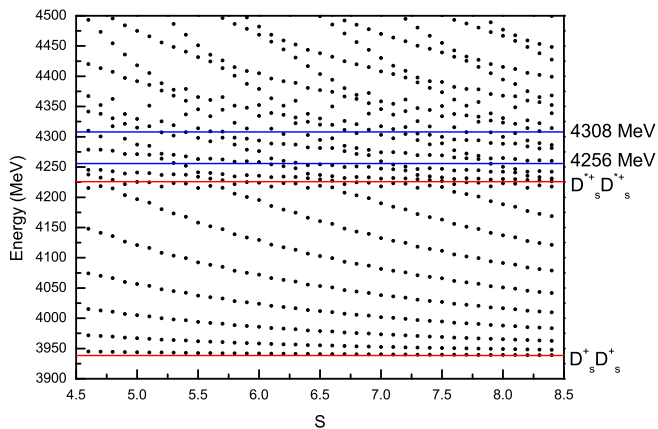} \vspace{-0.3in}

\caption{The stabilization plots of the energies of the $cc\bar{s}\bar{s}$ with $IJ^{P}=00^{+}$.}
\end{center}
\end{figure}

\begin{figure}[ht]
\begin{center}
\epsfxsize=3.14in \epsfbox{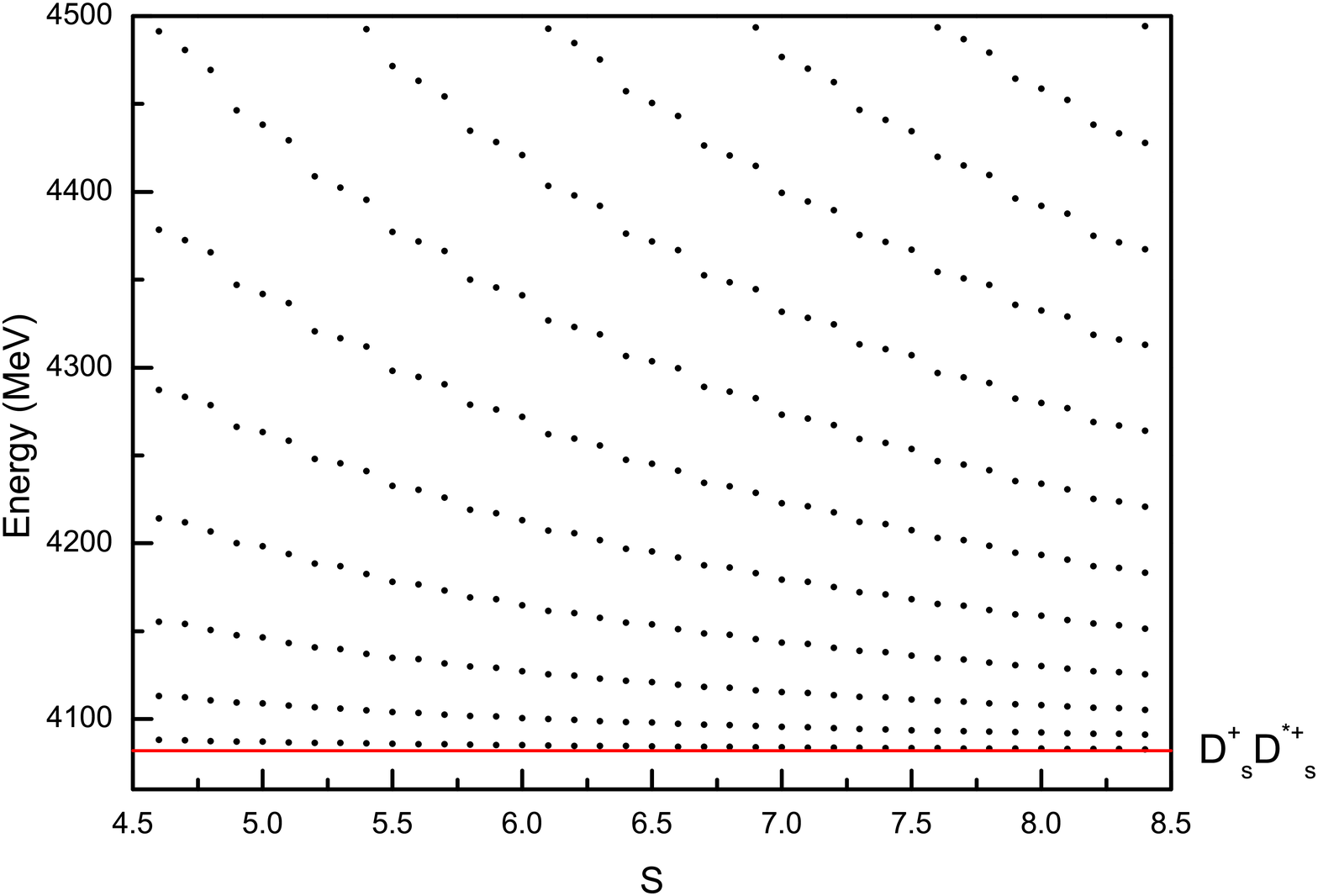} \vspace{-0.1in}

\caption{ The stabilization plots of the energies of the $cc\bar{s}\bar{s}$ with $IJ^{P}=01^{+}$.}
\end{center}
\end{figure}

\begin{figure}[ht]
\begin{center}
\epsfxsize=3.4in \epsfbox{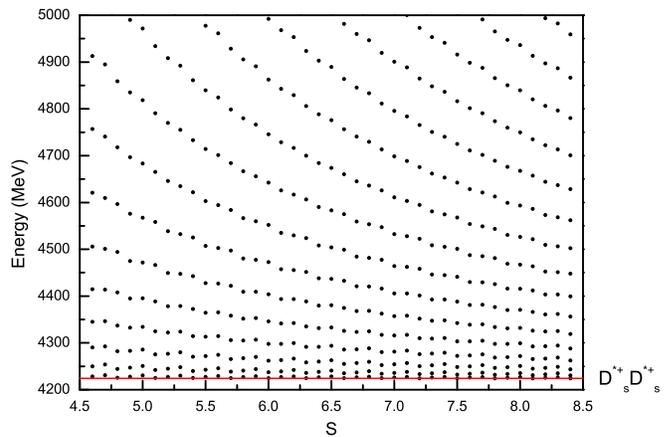} \vspace{-0.3in}

\caption{ The stabilization plots of the energies of the $cc\bar{s}\bar{s}$ with $IJ^{P}=02^{+}$.}
\end{center}
\end{figure}

\section{summary}
In this work, we systematically investigate the low-lying double-charm and double-strange tetraquark systems in the framework of the QDCSM. Two structures, meson-meson and diquark-antidiquark, as well as the coupling of these two configurations are considered.
The dynamical bound-state calculation is carried out to search for any bound state in the $cc\bar{s}\bar{s}$ systems. Besides, both the single channel and the channel coupling calculation are performed to investigate the effect of the channel coupling. Meanwhile, a stabilization calculation is carried out to find any resonance state.

The bound-state calculation shows that there is no any bound state for the $cc\bar{s}\bar{s}$ system in QDCSM. However, two resonance states with $IJ^{P}=0 0^{+}$ are obtained. One is more possible to be a molecular resonance state with a mass and width around $4256$ MeV and $60$ MeV, respectively; another one is more likely to be a compact resonance state with a mass and width around $4308$ MeV and $19$ MeV, respectively. Our results show that the coupling calculation is indispensable to explore the resonance states. In the work of the chiral quark model~\cite{Yanggang11}, there was no any bound state for the $cc\bar{s}\bar{s}$ system. However, several resonance states were obtained for $cc\bar{s}\bar{s}$ tetraquarks, which were one $IJ^{P}=0 0^{+}$ state with the resonance mass around $4.9$ GeV and three $IJ^{P}=02^{+}$ states with the resonance mass around $4.8$ GeV. Besides, the work of Ref.~\cite{R. Molina} also studied the state with $C=2$, $S=2$, $I=0$, and $J=0,~2$, and no any state was obtained in this sector. However, only the $D^{*}_{s}D^{*}_{s}$ channel was studied there. The coupling with other channels is worthy of consideration to find some resonance states.

Besides, although no significant signals were observed in present experiment at the Belle collaboration~\cite{Belle2021}, there is still some structures around $4.3$ GeV in the distributions of $M_{D^{+}_{s}D^{+}_{s}}$ and $M_{D^{*+}_{s}D^{*+}_{s}}$. We suggest that the experiment check with a larger amount of data in the future.

\acknowledgments{This work is supported partly by the National Science Foundation
of China under Contract Nos. 11675080, 11775118 and 11535005.

\end{document}